\documentclass[osajnl,preprint,showpacs,superscriptaddress,12pt]{revtex4-1} 
\usepackage{graphicx}
\usepackage{amsmath,amssymb,amsfonts}       

\usepackage{multirow}
\usepackage{color}
\usepackage{hyperref}%

\newcommand{\TFd}{\mathop{\mathbf{TF}}}

\newcommand{\average}[1]{\left \langle {#1} \right \rangle}

\newcommand{\lemmaend}{\hfill \ensuremath{\Box}}

\newcommand{\dint}{\mathrm{d}} 

\newcommand{\svec}{{\boldsymbol{s}}}

  \newcommand{\FTR}{\widetilde{\mathcal{R}}} 
  \newcommand{\G}{\mathcal{G}} 
  \newcommand{\FTG}{\widetilde{\mathcal{G}}} 
  \newcommand{\FTP}{\widetilde{\Pi}} 
  \newcommand{\Sl}{\mathbf{s}} 
  \newcommand{\FTS}{\widetilde{\mathbf{s}}} 
  \newcommand{\WF}{\boldsymbol{\varphi}} 
  \newcommand{\FTWF}{\widetilde{\boldsymbol{\varphi}}} 
  \newcommand{\Noise}{\boldsymbol{\eta}} 
  \newcommand{\FTN}{\widetilde{\boldsymbol{\eta}}} 

  \newcommand{\Sphi}{\mathbf{W}_\varphi} 
  \newcommand{\Ss}{\mathbf{W}_\mathbf{s}} 
  \newcommand{\Sn}{\mathbf{W}_\eta} 

\newenvironment{proof}[1][Proof]{\begin{trivlist}
\item[\hskip \labelsep {\bfseries #1}]}{\end{trivlist}}

\newcommand{\qed}{\nobreak \ifvmode \relax \else
      \ifdim\lastskip<1.5em \hskip-\lastskip
      \hskip1.5em plus0em minus0.5em \fi \nobreak
      \vrule height0.75em width0.5em depth0.25em\fi}

\begin{document}



\title{Anti-aliasing Wiener filtering for wave-front reconstruction in the spatial-frequency domain for high-order astronomical adaptive-optics systems}
\author{Carlos M. Correia}\email{ccorreia_at_astro.up.pt} 
\affiliation{Institute of Astrophysics and Space Sciences, University of Porto, CAUP, Rua das Estrelas, PT4150-762 Porto, Portugal}
\affiliation{Dept. of Physics and Astronomy, Faculty of Sciences, University of Porto,
Rua do Campo Alegre 687, PT4169-007 Porto, Portugal}
\affiliation{Adaptive Optics Laboratory, University of Victoria, 3800 Finnerty Rd., Victoria, BC, Canada V8P 5C2}
\author{Joel Teixeira$^{1,}$}
\affiliation{Dept. of Mathematics, Faculty of Sciences, University of Porto, Rua do Campo Alegre 687, PT4169-007 Porto, Portugal}


\begin{abstract}
Computationally-efficient wave-front reconstruction techniques for astronomical adaptive optics systems have seen a great development in the past decade. Algorithms developed in the spatial-frequency (Fourier) domain have gathered large attention specially for high-contrast imaging systems. 

In this paper we present the Wiener filter (resulting in the maximization of the Strehl-ratio) and further develop formulae for the anti-aliasing Wiener filter that optimally takes into account high-order wave-front terms folded in-band during the sensing (\textit{i.e.} discrete sampling) process. 

We employ a continuous spatial-frequency representation for the forward measurement operators and derive the Wiener filter when aliasing is explicitly taken into account. We further investigate and compare to classical estimates using least-squares filters the reconstructed wave-front, measurement noise and aliasing propagation coefficients as a function of the system order. Regarding high-contrast systems, we provide achievable performance results as a function of an ensemble of forward models for the Shack-Hartmann wave-front sensor (using sparse and non-sparse representations) and compute point-spread function raw intensities. 

We find that for a 32x32 single-conjugated adaptive optics system the aliasing propagation coefficient is roughly 60\% of the least-squares filters whereas the noise propagation is around 80\%.  Contrast improvements of factors of up to 2 are achievable across the field in H-band. For current and next generation high-contrast imagers, despite better aliasing mitigation, anti-aliasing Wiener filtering cannot be used as a stand-alone method and must therefore be used in combination with optical spatial filters deployed before image formation takes actual place.
\end{abstract}

\maketitle 

\section{Introduction}
In recent years the demand for high-contrast imaging (HCI) and tomographic systems with high-order adaptive-optics (AO) correction led to the development of computationally-efficient wave-front reconstruction (WFR) techniques in the spatial-frequency domain \cite{ellerbroek02,thiebaut10, gavel04}. Two main reasons can be pointed out: 

$i)$ the number of degrees-of-freedom became prohibitive with a computational burden scaling with $D^4$, i.e. the fourth power of the telescope diameter $D$ assuming a constant density of sensing/controlled points; this figure can be significantly relaxed to $D^2 log(D)$ with Fast-Fourier-Transform (FFT) techniques \cite{poyneer05, correia09,correia08} and 

$ii)$ the frequency modes can -- to a good degree of approximation -- be considered independent, even on a finite aperture \cite{poyneer07}, which fits well into modal gain optimization with predictive WF estimation for enhanced performance. Also, albeit to a lesser degree, Fourier modes permit the control of specific locations on the point-spread function (PSF).

The optimization of such methods has been given large attention in the past, both the spatial and temporal (used for prediction) components \cite{poyneer05, poyneer07}. However the aliasing error on the measurement model has never received specific treatment in the WFR process. Its propagation through the reconstruction is considered an important factor limiting the achievable contrast in high-contrast imaging (HCI) systems as SPHERE \cite{beuzit06, mouillet09}, GPI \cite{macintosh06} ScExAO \cite{martinache09} and PALM-3000 \cite{dekany13}.
The commonly used approach followed by all these instruments is to mitigate aliasing before the measurements are produced using the the spatially-filtered (SF) Shack-Hartmann (SH) wave-front sensor (WFS) \cite{poyneer04a}. 

Here instead we investigate and assess analytically a \textit{post-facto} approach, \textit{i.e.} we formulate the optimal, Strehl-ratio maximizing reconstructor with measurements affected by aliasing noise when no spatial filter is used. Although image quality cannot be fully recovered with \textit{post-facto} techniques (were it the case AO could be circumvented altogether in favor of deconvolution techniques~\cite{ellerbroek09}) we focus on the optimization of the reconstruction in the continuous spatial-frequency domain. 

We present both the least-squares (LSQ) and minimum-mean-squared-error (MMSE) Wiener filters and further develop formulae for the anti-aliasing (AA) Wiener filter. The latter is complementary to the SF-SH-WFS and can be used as a reference standard that achieves minimum residual WF error variance (and therefore maximum Strehl-ratio) against which sub-optimal methods can be compared. The noise propagation and aliasing estimates presented in \textit{Ellerbroek et al}~\cite{ellerbroek05} for the LSQ case are updated for the spatial Wiener filters. 
We provide wave-front error figures and an error breakdown for noise and aliasing. Furthermore, we provide achievable Strehl-ratio estimates and contrast improvement factors as a function of the system order. While analytical tools developed herein cannot be used for detailed system design they can be of great help in quickly exploring vast swaths of parameters that are crucial for the design. Such analytical tools follow the examples of \textit{Cibola}~\cite{ellerbroek05}, \textit{PAOLA}~\cite{jolissaint10} and more recently for laser tomographic systems \cite{tatulli13} to site a few. 

High-contrast imaging (HCI) systems offer a suitable and attractive scenario of application and it is believed real-time spatial-frequency reconstructors can provide optimal or near-optimal performance -- take the case of GPI \cite{poyneer04a}. Iterative extensions to the tomographic case have been pursued \cite{gavel04, yang05} as well as the Linear-Quadratic-Gaussian in the spatial-frequency domain \cite{correia09}. 

Targeting the second generation high-contrast imagers, we compare results to those obtained using approximate measurement models for the SH-WFS that admit sparse representations, the latter considered preferred for real-time implementation. We adopt a continuous representation that can be later straightforwardly translated to the discrete case in line with previous work in \cite{correia08} and \cite{poyneer07} for real-time application.
  
This document is organized as follows: section 2 presents the SH-WFS model and gives formulae for the spatial-frequency domain representation; in section 3 the LSQ and Wiener filters are derived. WF errors are investigated in section 4 whereas section 5 gives sample examples of power-spectral densities and PSF raw intensities for high-order systems.

  \section{SH-WFS forward model}
  
  We start our analysis by describing the forward model in the space domain. The spatial-frequency domain model is provided afterwards along with discrete first differences approximate models common in the literature and extensively used in AO modeling. 
  
  A common working assumption is that measurements are open-loop, an oversimplification that allows us to provide optimal filters which involve known, stationary phase and noise statistics \cite{ellerbroek05, poyneer03, rigaut98, flicker07a}. 

Adaptation to the negative-feedback closed-loop case as well as assessment of non-Gaussian-distributed additive measurement noise is left for a subsequent paper with a pseudo-open-loop framework likely to be used \cite{gilles05, piatrou05}.

In the remainder we follow closely the notation from \textit{Rigaut et al} \cite{rigaut98} kept later in \textit{Flicker et al}~\cite{flicker07a} and assume the Fried geometry \cite{fried77} whereby the DM actuators are placed at the corners of the SH-WFS sub-apertures -- a configuration employed both on SPHERE, GPI and other high-order AO systems. There are however a few exceptions namely AO systems with deformable secondaries which for opto-mechanical reasons tend to follow a radial arrangement of the actuators. 

  \subsection{Space domain}
  
  Let the SH-WFS measurements $\Sl(\mathbf{x},t)$ be given by the
  geometrical-optics linear model \cite{roddier99, rigaut98, hardy98}   
  \begin{equation}\label{eq:s_Gphi}
    \Sl(\mathbf{x},t) = \G\WF(\mathbf{x},t) + \Noise(\mathbf{x},t),
  \end{equation}
  where $\G$ is a phase-to-slopes linear operator mapping
  aperture-plane guide-star
  wave-fronts $\WF(\mathbf{x},t)$ into WFS measurements over a bi-dimensional space indexed by $\mathbf{x} = (x,y)$ at time $t$;  $\Noise(\mathbf{x},t)$ represents white
  noise due to photon statistics,
  detector read noise and background photons. Both $\WF$ and $\Noise$ are zero-mean functions of Gaussian probability distributions and known covariance matrices $\Sigma_\phi$ and
  $\Sigma_\eta$ respectively. Noise is assumed both temporally and spatially uncorrelated.

  For the SH-WFS, a 2D map of slopes is obtained from the discretization of the average gradient over the lenslets conjugated to the pupil-plane
  \begin{equation}\label{eq:theoretical_measurement_partialderivative}
    \svec[m,n] = 
    \frac{1}{d}\left[\int_{x_n}^{x_{n+1}}\int_{y_m}^{y_{m+1}}
      \frac{1}{T_s}\int_{-T_s/2}^{T_s/2} \nabla \WF(x,y,t) \partial t \partial y \partial x\right],
  \end{equation}
  where $d$ is the sub-aperture width in meters, $x_{n+1} = x_n + d$, $y_{m+1} = y_m + d$ are the integration bounds corresponding to the sub-aperture edges and $\svec[m,n] \triangleq \svec(m d, n d)$, $(m,n) \in \mathbb{Z}$ are the discrete measurements sampled at the corners of 
  the sub-apertures.
The integral over time models temporal integration on the sensor over a $T_s$ period; no extra delay in the loop is considered although this could be included with minimal effort. 
  The gradient operator $\nabla$ is defined as
  \begin{equation}
    \nabla  \triangleq \left(\begin{array}{c}
        \frac{\partial }{\partial x} \\
        \frac{\partial }{\partial y}
      \end{array}
    \right),
  \end{equation}
  and the discretization process can be exactly modeled as a multiplication of the continuous slopes by a comb function
  \begin{equation}
    \mathsf{III}\left(\frac{\mathbf x}{d}\right) \triangleq \sum_{m=-\infty}^{\infty}
    \sum_{n=-\infty}^{\infty} \delta\left(\frac{x}{d}-m\right)\delta\left(\frac{y}{d}-n\right) .
  \end{equation}
with $\delta (\cdot)$ the Dirac ``delta'' function.

  The average wave-front is determined from the
  convolution by a squared function of width $d$. Since samples
  are to be taken at sub-aperture's corners, a
  half-sub-aperture shift is introduced in the squared function. For the temporal integration we used the frozen flow hypothesis, i.e. $\WF(\mathbf{x},t+\tau) = \WF(\mathbf{x} - \mathbf{v}\tau,t)$ with $\mathbf{v} = (v_x, v_y)$ the wind velocity vector.  

Putting this all together one gets
  \begin{equation}
    \G = \mathsf{III}\left(\frac{\mathbf{x}}{d}\right) \times \left[\Pi\left(\frac{\mathbf{x}-1/2}{d}\right) \otimes\Pi\left(\frac{\mathbf{x}}{\mathbf{v}T_s}\right) \otimes \nabla\right],
  \end{equation}
  where $\otimes$ is a 2-dimensional convolution product, $\times$ is point-wise multiplication and $\Pi (\cdot)$ 
  the "square" separable function 
  \begin{equation}
    \Pi(\mathbf{x})   \triangleq \left\{\begin{array}{ll}
        1&  \text{if}\,\, |x|\leq 1/2 \land |y| \leq 1/2 \\
        0& \text{otherwise}
      \end{array}
    \right. .
  \end{equation}

  The exact solution to
  Eq.(\ref{eq:theoretical_measurement_partialderivative})   can be determined from the difference of the
  average phase at the opposite edges of the sub-aperture
  (top/bottom and left/right edges, depending on whether the
  $x$ or $y$ direction is considered). By definition the average gradient is the difference of the extreme points. Thus, neglecting for a moment the temporal integration of the WF (note the spatial and temporal integrals inter-change),
  \begin{equation}\label{eq:theoretical_x_measurement}
    \svec_x[m,n] = 
    \frac{1}{d}\left[\int_{y_{m}}^{y_{m+1}}\WF(x_{n+1}, y)
      \partial y - \int_{y_{m}}^{y_{m+1}} \WF(x_n,y) \partial y\right],
  \end{equation}
  for the x-slopes and 
  \begin{equation}\label{eq:theoretical_y_measurement}
    \svec_y[m,n] 
    = 
    \frac{1}{d}\left[\int_{x_{n}}^{x_{n+1}}\WF(x, y_{m+1})
      \partial x -
      \int_{x_{n}}^{x_{n+1}}\WF(x,y_m) \partial x\right], 
  \end{equation}
  for the y-slopes. This formulation provides insight into the models that will be developed next, in particular the discrete, sparse approximations.


The complete SH model is dubbed 'Rigaut' since it inherits from initial work by Rigaut et al in \cite{rigaut98}. 
  \subsection{Continuous model in the spatial-frequency domain}

  Using the fact that the SH-WFS measurement model is a set of convolution integrals, treatment in the spatial-frequency domain is straightforward. 
  Let the Fourier-domain representation of Eq. (\ref{eq:s_Gphi})
  \begin{equation}\label{eq:S_FT}
    \FTS\left(\boldsymbol{\kappa}\right) = \FTG \FTWF\left(\boldsymbol{\kappa}\right) + \widetilde{\boldsymbol{\eta}}\left(\boldsymbol{\kappa}\right),
  \end{equation}
  with $\boldsymbol{\kappa} = (\kappa_x,\kappa_y) \in \mathbb{R}^2$ the frequency vector and symbol $\widetilde{\,\,\cdot\,\,}$ used for Fourier-transformed variables. The time-dependence is now dropped following the frozen-flow hypothesis. Using common transform pairs for the individual operations (see e.g.
  \cite{oppenheim97}), the Fourier
  representation of the measurements in 
  Eqs. (\ref{eq:theoretical_x_measurement}-\ref{eq:theoretical_y_measurement})
  results in
  \begin{equation}\label{eq:FT_G}
    \FTG = \TFd\{\G\} = \mathsf{III}\left(\boldsymbol{\kappa} d\right) \otimes [\FTP\left(\boldsymbol{\kappa} d\right) \times e^{i \pi d \boldsymbol{\kappa}} \times \widetilde{\nabla} \times \FTP\left(\boldsymbol{\kappa} \mathbf{v}T_s\right)],
  \end{equation}
  with $\TFd\{\cdot\}$ the continuous Fourier transform and $i\triangleq\sqrt{-1}$.  
From this last equation the Fourier transform pairs are
  \begin{equation}
    \widetilde{\nabla}(\boldsymbol{\kappa}) = 2 i \pi d \boldsymbol{\kappa} = 2 i \pi d [\kappa_x, \kappa_y],
  \end{equation}
and
  \begin{align}
    \FTP\left(\boldsymbol{\kappa} d\right)e^{i \pi d \boldsymbol{\kappa}} & = sinc (d\boldsymbol{\kappa}) \times e^{i \pi d \boldsymbol{\kappa}}\\  
    & = 
    \frac{sin(\pi \kappa_x d)}{\pi\kappa_x d} \times \frac{sin(\pi \kappa_y d)}{\pi \kappa_y d} \times e^{i \pi d (\kappa_x + \kappa_y)}  
  \end{align}
  where $sinc(x) \triangleq sin(\pi x)/ (\pi x)$ and the exponential term shifts the slopes by half a sub-aperture width. The temporal averaging function is promptly
  \begin{equation}
    \FTP\left(\boldsymbol{\kappa} \mathbf{v}T_s\right) = sinc\left(\boldsymbol{\kappa} \mathbf{v}T_s\right)
  \end{equation}
on account of the frozen-flow assumption made earlier.

  Equation (\ref{eq:FT_G}) admits also the following representation
  \begin{subequations}\label{eq:Gxy_FT}
    \begin{align}
      \FTG_x & = \left[\left(e^{2 i \pi d \kappa_x} -1 \right) \times sinc(\kappa_y d) e^{i \pi d \kappa_y} \times \FTP\left(\boldsymbol{\kappa} \mathbf{v}T_s\right) \right] \otimes \mathsf{III}\left(\boldsymbol{\kappa} d\right)\\
      \FTG_y & = \left[\left[e^{2 i \pi d \kappa_y} -1 \right) \times sinc(\kappa_x d) e^{i \pi d \kappa_x}  \times \FTP\left(\boldsymbol{\kappa} \mathbf{v}T_s\right)  \right]\otimes \mathsf{III}\left(\boldsymbol{\kappa} d\right)
    \end{align}
  \end{subequations}
  where the term within square brackets brings to light the discrete difference of phase at apposite edges of the sub-apertures; Not only it translates intuitively the nature of Eqs. (\ref{eq:theoretical_x_measurement}-\ref{eq:theoretical_y_measurement}) a straightforward parallel can be made with the approximate discrete gradient models proposed further below. The symbol $\times$ will be droped in the remainder.


  \begin{proof}
Take the $'x'$ direction in Eq. (\ref{eq:FT_G}).
Splitting the averaging term into its factors, the terms in $\kappa_{y}$ vanish and the exponential terms are aggregated from standard $sin(\cdot)$ and $cos(\cdot)$ functions as follows
\begin{align*}
 \FTP\left(\boldsymbol{\kappa} d\right)  e^{i \pi d \boldsymbol{\kappa}}  \widetilde{\nabla}_x
 & =  2i\pi \kappa_x d \frac{sin(\pi \kappa_x d)}{\pi\kappa_x d} \frac{sin(\pi \kappa_y d)}{\pi \kappa_y d} e^{i \pi d (\kappa_x + \kappa_y)} \\
  & = 2i\ sin(\pi\kappa_x d)\ sinc(\kappa_y d)\ e^{i \pi d \kappa_x}\ e^{i \pi d \kappa_y} \\
 & = (e^{2i \pi d \kappa_x} - 1)\ sinc(\kappa_y d)\ e^{i \pi d \kappa_y}
\end{align*}
and analogously for the $'y'$ direction.
  \lemmaend
  \end{proof}
In order to properly account for the spectral replication as a result of the convolution by the comb function -- in other words the aliasing term -- let us now split the WF according to the cut-off frequencies into a in-band and a out-of-band term: $\FTWF\left(\boldsymbol{\kappa}\right) = \FTWF_\parallel\left(\boldsymbol{\kappa}\right) + \FTWF_\perp\left(\boldsymbol{\kappa}\right)$ with 
  \begin{equation}
   \FTWF_\parallel\left(\boldsymbol{\kappa}\right) \triangleq \left\{\begin{array}{ll}
       \FTWF\left(\boldsymbol{\kappa}+ \mathbf{m}/d \right), & \text{if  } \mathbf{m} = 0
       \\
        0&  \text{if  } \mathbf{m} \neq 0 
      \end{array}
    \right. ,
  \end{equation}
in which $\mathbf{m} = (m,n), m,n \in \mathbb Z$ and $\FTWF_\perp\left(\boldsymbol{\kappa}\right)$ is the complement set of $\FTWF_\parallel\left(\boldsymbol{\kappa}\right)$.
With the above definition the measurement Eq. (\ref{eq:S_FT}) can be expanded to
  \begin{align}
    \FTS\left(\boldsymbol{\kappa}\right) & = 2 i \pi d \sum_{\mathbf{m}} \left(\boldsymbol{\kappa} + \mathbf{m}/d \right) \FTWF\left(\boldsymbol{\kappa} + \mathbf{m}/d \right)
    \FTP\left(\boldsymbol{\kappa}d + \mathbf{m}\right)e^{i \pi (\boldsymbol{\kappa}d + \mathbf{m})}\FTP\left[(\boldsymbol{\kappa} + \mathbf{m}/d) \mathbf{v}T_s\right] + \FTN\left(\boldsymbol{\kappa}\right) \nonumber
    \\ & =  2 i \pi d \boldsymbol{\kappa} \FTP\left(\boldsymbol{\kappa}d \right) e^{i \pi d \boldsymbol{\kappa}} \FTP\left(\boldsymbol{\kappa} \mathbf{v}T_s\right) \FTWF_\parallel\left(\boldsymbol{\kappa} \right)  \nonumber \\ & + \underbrace{ 2 i \pi d \sum_{\mathbf{m}\neq 0} \left(\boldsymbol{\kappa} + \mathbf{m}/d \right) \FTWF\left(\boldsymbol{\kappa} + \mathbf{m}/d \right)
    \FTP\left(\boldsymbol{\kappa}d + \mathbf{m}\right)e^{i \pi (\boldsymbol{\kappa}d + \mathbf{m})}\FTP\left[(\boldsymbol{\kappa} + \mathbf{m}/d) \mathbf{v}T_s\right] + \FTN\left(\boldsymbol{\kappa}\right)}_{\text{Generalised Measurement Noise}} \label{eq:S_FT_AliasTerm}
  \end{align}
where it becomes apparent that the wave-front sensing operator $\FTG$ is not purely a spatial filter except for functions within the pass-band. The  term over the under-brace acts as a type of generalized measurement noise, composed of straight photon shot and electron noise plus spatial aliasing of high-order WF terms folded in-band during the sampling process \cite{ellerbroek05}. This feature will allow us to proceed and synthesize filters to optimally estimate the wave-front within the correctable band.

  \subsection{Approximate discrete measurement models}\label{sec:approx_disc_meas_models}
\begin{figure}[htpb]
    \begin{center}
      \includegraphics[width=1.0\textwidth]{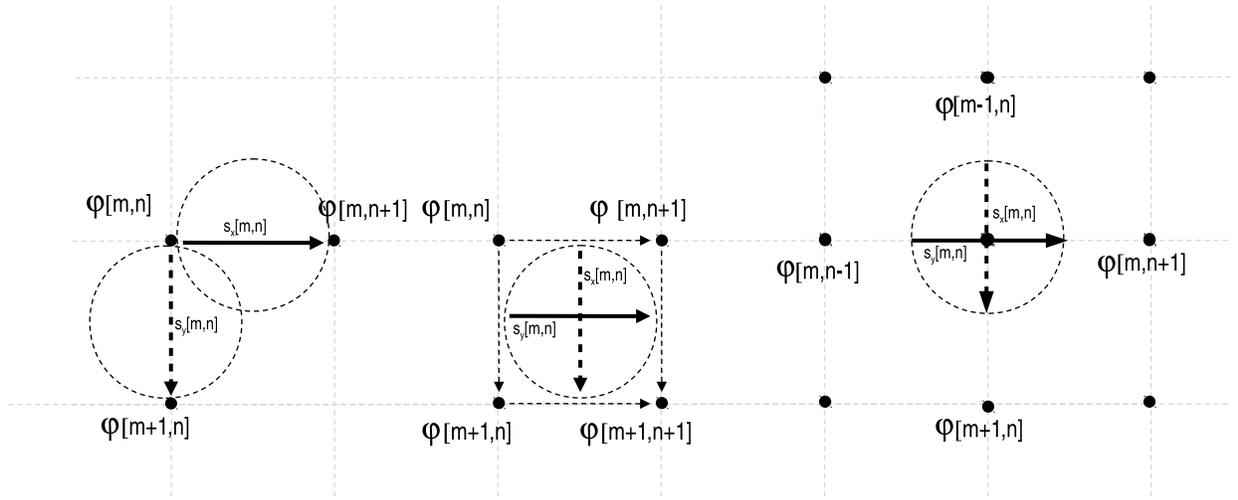} 
    \end{center}
    \caption[]
    {\label{fig:Fried_Hudgin_Geometry}
      Hudgin (left),  Fried (center) and Southwell (right) models for the SH-WFS. Circles represent the phase points required for the slope measurement for each model.}
  \end{figure}

  To represent the measurement model of
  Eqs. (\ref{eq:theoretical_measurement_partialderivative}) and ~(\ref{eq:theoretical_x_measurement}-\ref{eq:theoretical_y_measurement}) various sensor models 
  have been proposed \cite{freischlad86}, which are briefly recalled in Fig.~(\ref{fig:Fried_Hudgin_Geometry}). They
  consist in discrete first differences approximations to the spatially-averaged gradient output by the SH-WFS which represent in more or less detail the measurement it provides -- note that the temporal integration is neglected altogether. Unlike the model presented in the previous section, these models have a sparse
  representation in the direct domain and have been extensively
  used both in iterative and non-iterative sparse methods \cite{yang05, ellerbroek03,
    gilles02, thiebaut10}. 


The so-called Hudgin \cite{hudgin77} geometry
  (Fig.~(\ref{fig:Fried_Hudgin_Geometry}), left) assumes SH
  measurements are the discrete first differences between two
  adjacent phase points.
  This simple model neglects the wave-front averaging over the edges of the sub-apertures. 
  In order to increase the match between the models and the actual SH-WFS measurements it has been previously suggested that an extra alignment term of $1/8\,d$ can be introduced on the forward model \cite{poyneer03}. With this extra term, the continuous spatial frequency domain representation becomes
   \begin{subequations}\label{eq:Hudgin_slopes_FT}
    \begin{align}
      \FTS_x^H\left(\boldsymbol{\kappa}\right) & = e^{1/4\,i\pi d\left(\kappa_x+\kappa_y\right)} \left(e^{2 i \pi d \kappa_x} -1 \right) \FTWF(\boldsymbol{\kappa})\\
      \FTS_y^H\left(\boldsymbol{\kappa}\right) & = e^{1/4\,i\pi d\left(\kappa_x+\kappa_y\right)} \left(e^{2 i \pi d \kappa_y} -1 \right) \FTWF(\boldsymbol{\kappa}).
    \end{align}
  \end{subequations}



  The Fried model 
  (Fig.~(\ref{fig:Fried_Hudgin_Geometry}), center) considers a
  two-point averaging over opposite edges of the sub-aperture \cite{fried77}
  Fried's model implicitly
  assumes the expansion of the phase $\WF$ on the basis of
  bi-linear splines. In other words, the Fried geometry is exact
  (with respect to Eqs.(\ref{eq:theoretical_x_measurement}) and
  (\ref{eq:theoretical_y_measurement})) should the phase be
  expanded onto a basis of bi-linear splines in the noiseless
  case. Its spatial-frequency domain representation is therefore
   \begin{subequations}\label{eq:Fried_slopes_FT}
    \begin{align}
      \FTS_x^F\left(\boldsymbol{\kappa}\right) & = 1/2 \left(e^{2 i \pi d \kappa_x} -1 \right) \left(e^{2 i \pi d \kappa_y} + 1  \right) \FTWF(\boldsymbol{\kappa}) \\
      \FTS_y^F\left(\boldsymbol{\kappa}\right) & = 1/2 \left(e^{2 i \pi d \kappa_y} -1 \right) \left(e^{2 i \pi d \kappa_x} + 1  \right) \FTWF(\boldsymbol{\kappa}).
    \end{align}
  \end{subequations}

  Finally the Southwell geometry
  (Fig.~(\ref{fig:Fried_Hudgin_Geometry}), right) assumes
  measurements are located at the intersections of the grid~\cite{southwell80},
  exactly where the phase points are to be reconstructed. 
It translates into the spatial-frequency domain as \cite{correia08}
  \begin{subequations}\label{eq:Southwell_slopes_FT}
    \begin{align}
      \FTS_x^S\left(\boldsymbol{\kappa}\right)\left(e^{2 i \pi d \kappa_x} + 1\right)/2 
      & = e^{i\pi d\left( \kappa_x+\kappa_y\right)}\left(e^{2 i \pi d \kappa_x} -1 \right) \FTWF(\boldsymbol{\kappa})  \\
      \FTS_y^S\left(\boldsymbol{\kappa}\right)\left(e^{2 i \pi d \kappa_y} + 1 \right)/2
      & = e^{i\pi d\left( \kappa_x+\kappa_y\right)}\left(e^{2 i \pi d \kappa_y} -1 \right) \FTWF(\boldsymbol{\kappa}),
    \end{align}
  \end{subequations}
where we added a half sub-aperture shift found after numerical simulations following the same reasoning applied in the Hudgin case. 



  \section{Wave-front reconstruction: filter synthesis techniques}

Wave-front reconstruction filters are now synthesized in order to estimate (reconstruct) the wave-fronts from  WFS measurements. Both least-squares and \textit{minimum-mean-square-error} (MMSE) minimization criteria are utilized. The latter is also called Wiener filter \cite{wiener49} and such label will be kept throughout this paper. It is the spatial-frequency equivalent to Strehl-optimal estimators derived directly in the spatial domain \cite{wallner83, ellerbroek05, ellerbroek09}.

We proceed by neglecting the aliased component on the forward measurement model and later explicitly incorporating it to develop the Anti-Aliasing Wiener filter.
  \subsection{Least-squares reconstruction filters}
  Let the least-squares cost-functional
  \begin{equation}\label{eq:LSQcrit}
    \widehat{\FTWF}(\boldsymbol{\kappa}) = 
    \arg \min_{\FTWF(\boldsymbol{\kappa})}
    \left|\FTS - \FTG \FTWF(\boldsymbol{\kappa}) \right|^2,
  \end{equation}
  where one attempts to finding the wave-front that best fits to the measured slopes. Estimated variables use a \textit{hat} symbol overhead. This practical estimator is not related to any imaging assumption nor statistical knowledge of the stationary processes involved. However for high SNR regimes its use is perfectly justified. 

Historically LSQ has been adopted since an empirical model for $\G$ can be measured on an optical bench when the phase is expressed in the DM influence functions. Solving for the DM commands using truncated singular-value decomposition was therefore (and still is) a very practical and effective solution. To the author's knowledge every classical AO system around the globe uses this approach. The reason why it may be not sufficient for more advanced systems is that the very high performance levels requested will make optimization through regularization and advanced filtering a prerequisite namely for HCI systems.

The solution to Eq. (\ref{eq:LSQcrit}) is readily given by
\begin{equation}
\FTWF(\boldsymbol{\kappa})  = \frac{1}{\FTG} \FTS = \frac{\FTG^*}{\left|\FTG\right|^2} \FTS = \frac{\FTG_x^* \FTS_x(\boldsymbol{\kappa}) +\FTG_y^* \FTS_y(\boldsymbol{\kappa})}{\left|\FTG_x\right|^2 +\left|\FTG_y\right|^2},
\end{equation}
which isn't but the immediate inverse of the forward measurement model \cite{freischlad86, correia08, poyneer03}.
 
Using the properties of the Fourier Transform and solving for the phase in
  Eqs. (\ref{eq:Hudgin_slopes_FT}), Eq. (\ref{eq:Fried_slopes_FT}) and Eq.~(\ref{eq:Southwell_slopes_FT}), the filter $\FTR = (\mathcal
  \FTR_x, \FTR_y)$  with $\widehat{\FTWF}(\boldsymbol{\kappa})= \FTR_x\FTS_x(\boldsymbol{\kappa}) + \FTR_y\FTS_y(\boldsymbol{\kappa})$ is 
 respectively for the [Fried | Hudgin | Southwell]  model-based filters 

\begin{subequations}
 \begin{align}\label{eq:filtro_fried}
      \FTWF(\boldsymbol{\kappa}) & =  \Big[\left(e^{-2 i \pi d \kappa_x} -1 \right) \left(e^{-2 i \pi d \kappa_y} + 1  \right) \,\,\FTS_x^F\left(\boldsymbol{\kappa}\right)
 + \left(e^{-2 i \pi d \kappa_y} -1 \right) \left(e^{-2 i \pi d \kappa_x} + 1  \right)\,\,\FTS_y^F\left(\boldsymbol{\kappa}\right) \Big] \nonumber \\
& \hspace{50pt}1/8 \Big[ \sin^2(\pi d \kappa_x) \cos^2(\pi d \kappa_y) + \sin^2(\pi d \kappa_y)\cos^2(\pi d \kappa_x) \Big]^{-1}\,\, ,\\&\nonumber\\ 
\FTWF(\boldsymbol{\kappa}) & = e^{-1/4 i\pi d\left(\kappa_x+\kappa_y\right)} \left\{ \frac{ \left(e^{-2 i \pi d \kappa_x} -1 \right) }{4 \Big[ \sin^2(\pi d \kappa_x) + \sin^2(\pi d \kappa_y)\Big]} \FTS_x^H\left(\boldsymbol{\kappa}\right) \right. + \nonumber\\
   & \hspace{100pt} \left. \frac{\left(e^{-2 i \pi d \kappa_y} -1 \right)}{4 \Big[ \sin^2(\pi d \kappa_x) + \sin^2(\pi d \kappa_y)\Big]}\FTS_y^H\left(\boldsymbol{\kappa}\right)\right\},\\&\nonumber\\
\FTWF(\boldsymbol{\kappa}) & = e^{-i\pi d\left(\kappa_x+\kappa_y\right)} \left\{\frac{-i\,\sin (2\pi d \kappa_x)}{4 \Big[ \sin^2(\pi d \kappa_x) + \sin^2(\pi d \kappa_y)\Big]}\FTS_x^S\left(\boldsymbol{\kappa}\right)  + \right.\nonumber \\
 & \hspace{100pt} \left. \frac{- i\, \sin (2\pi d \kappa_y)}{4 \Big[ \sin^2(\pi d \kappa_x) + \sin^2(\pi d \kappa_y)\Big]}\FTS_y^S\left(\boldsymbol{\kappa}\right)\right\}.
 \end{align}

 \end{subequations}
  From the above equations, it can be easily checked that all the filters are of the form
\begin{subequations}
     \begin{align}\label{eq:Filter_x}
      \widetilde{\mathcal R}_x(\boldsymbol{\kappa}) & = 
      \frac{ \widetilde{\mathcal{D}}_x^*\widetilde{\mathcal{A}}_x^*\widetilde{\mathcal{S}}_x^*}{|\widetilde{\mathcal{D}}_x\widetilde{\mathcal{A}}_x|^2 +  |\widetilde{\mathcal{D}}_y \widetilde{\mathcal{A}}_y|^2} \widetilde{\mathcal{E}},\\
       \widetilde{\mathcal R}_y(\boldsymbol{\kappa}) & = 
      \frac{ \widetilde{\mathcal{D}}_y^*\widetilde{\mathcal{A}}_y^*\widetilde{\mathcal{S}}_y^*}{|\widetilde{\mathcal{D}}_x\widetilde{\mathcal{A}}_x|^2 +  |\widetilde{\mathcal{D}}_y \widetilde{\mathcal{A}}_y|^2}\widetilde{\mathcal{E}},
    \end{align}
  \end{subequations}
where 
 $\widetilde{\mathcal D}(\boldsymbol{\kappa})$ is the purely gradient-taking part, 
  $\widetilde{\mathcal A}(\boldsymbol{\kappa})$ and $\widetilde{\mathcal S}(\boldsymbol{\kappa})$ are averaging functions on the phase and slopes respectively, depending on the filter in
  use and $\widetilde{\mathcal E}(\boldsymbol{\kappa})$ are spatial shifts to increase the correlation between the model and the SH-WFS. Table~\ref{tab:Filters}  summarizes the filters
  considered before.

  \begin{table}[htpb]
    \caption[Resumo dos componentes dos filtros de reconstrução.] {\label{tab:Filters}
      Review of the filters for the Hudgin, Fried, Southwell
      geometries. 
}
    \begin{center}
      \begin{tabular}{cc|c|c|c}

 & \multicolumn{4}{ c }{\textbf{Filter}}\\
 \cline{2-5}
 \multicolumn{1}{ c| }{ }& Rigaut  & Fried & Hudgin & \multicolumn{1}{ c| }{Southwell }\\
 \hline\hline
  \multicolumn{1}{ |c| }{$\widetilde{\mathcal{D}}_x$}
  & $e^{2i\pi d\kappa_x}-1$
  & $e^{2i\pi d\kappa_x}-1$
  & $e^{2i\pi d\kappa_x}-1$
  & \multicolumn{1}{ c| }{$e^{2i\pi d\kappa_x}-1$}
  \\
  \hline
  \multicolumn{1}{ |c| }{$\widetilde{\mathcal{D}}_y$}
  & $e^{2i\pi d\kappa_y}-1$
  & $e^{2i\pi d\kappa_y}-1$
  & $e^{2i\pi d\kappa_y}-1$
  & \multicolumn{1}{ c| }{ $e^{2i\pi d\kappa_y}-1$}
  \\
  \hline
  \multicolumn{1}{ |c| }{ }& \multicolumn{1}{ c| }{\multirow{2}{*}{$sinc\left(d \kappa_y\right)e^{i \pi d \kappa_y }$} }&&&\multicolumn{1}{ c| }{\multirow{3}{*}{}}\\
  \multicolumn{1}{ |c| }{$\widetilde{\mathcal{A}}_x$}
  & \multicolumn{1}{ c| }{\multirow{2}{*}{$sinc\left(\boldsymbol{\kappa v}T_s\right)$} }
  & $\frac{1}{2}\left(1+e^{2i\pi d \kappa_y} \right)$
  & 1
  & \multicolumn{1}{ c| }{ 1}
  \\
  \multicolumn{1}{ |c| }{ }&&&&\multicolumn{1}{ c| }{ }\\
  \hline
  \multicolumn{1}{ |c| }{ }& \multicolumn{1}{ c| }{\multirow{2}{*}{$sinc\left(d \kappa_x\right)e^{i \pi d \kappa_x }$} }&&&\multicolumn{1}{ c| }{\multirow{3}{*}{}}\\
  \multicolumn{1}{ |c| }{$\widetilde{\mathcal{A}}_y$}
  & \multicolumn{1}{ c| }{\multirow{2}{*}{$sinc\left(\boldsymbol{\kappa v}T_s\right)$} }
  & $\frac{1}{2}\left(1+e^{2i\pi d \kappa_x} \right)$
  & 1
  & \multicolumn{1}{ c| }{ 1}
  \\
  \multicolumn{1}{ |c| }{ }&&&&\multicolumn{1}{ c| }{ }\\
  \hline
  \multicolumn{1}{ |c| }{$\widetilde{\mathcal{S}}_x^\ast$}
  & 1
  & 1
  & 1
  & \multicolumn{1}{ c| }{ $\frac{1}{2}\left(1+e^{2i\pi d \kappa_x} \right)$}
  \\
  \hline
  \multicolumn{1}{ |c| }{$\widetilde{\mathcal{S}}_y^\ast$}
  & 1
  & 1
  & 1
  &\multicolumn{1}{ c| }{ $\frac{1}{2}\left(1+e^{2i\pi d \kappa_y } \right)$}
  \\
  \hline
  \multicolumn{1}{ |c| }{$\widetilde{\mathcal{E}}$}
  & 1
  & 1
  & $e^{-1/4\,i\pi d\left(\kappa_x+\kappa_y\right)}$
  & \multicolumn{1}{ c| }{ $e^{-i\pi d\left(\kappa_x+\kappa_y\right)}$}
  \\
        \hline\hline
      \end{tabular}
    \end{center}
  \end{table}

It has been recognized previously that the Fried geometry is particularly affected at the waffle frequency $|\boldsymbol{\kappa}| =\pm 1/(2d)$.
Although not explicitly included in the summary table, further filtering can be implemented following the treatment in Poyneer \textit{et al} \cite{poyneer03} whereby the waffle-removal filter is
\begin{equation}
\widetilde{\mathcal{W}}(\boldsymbol{\kappa}) = \frac{1}{4}\Big[ 3 + e^{-2\pi i d \kappa_y} + e^{-2 \pi i d \kappa_x} - e^{-2 \pi i d (\kappa_x +\kappa_y)} \Big]. 
\end{equation}

  \subsection{Wiener filtering reconstruction}
  Instead of the best-fit to the measurements, let's now focus on minimizing the residual variance within the correctable band $|\boldsymbol{\kappa}|\leq 1/(2d)$ which is equivalent to maximizing the Strehl-ratio (the ratio of peak intensity of the aberrated point-spread-function (PSF) to the diffraction-limited PSF)
  \begin{equation}\label{eq:WienerFilterMinimFunctional}
    \FTR_\text{Wiener} 
    = \arg \min_{\FTR}\average{\boldsymbol{\varepsilon}^2(\boldsymbol{\kappa})} = \arg \min_{\FTR}\average{\left| \FTWF(\boldsymbol{\kappa})-\FTR\FTS(\boldsymbol{\kappa})  \right|^2}, 
  \end{equation}
  with $\widehat{\FTWF}(\boldsymbol{\kappa}) = \FTR_\text{Wiener}\FTS(\boldsymbol{\kappa})$ 
  and $\average{\cdot}$ stands for ensemble averaging over the turbulence and noise statistics.

  In the spatial-frequency domain, the solution to this minimization problem is given by the
  minimum mean-square error (MMSE) Wiener filter (assuming the
  signal and noise processes 
  are second-order stationary)~\cite{wiener49} 
  \begin{align}\label{eq:Phi_wiener}
    \FTR_{\text{\tiny{Wiener}}}(\boldsymbol{\kappa}) & = \frac{\FTG^*}{|\FTG|^2 + \gamma \frac{\Sn}{\Sphi}} = \frac{{\FTG}^* \Sphi}{|{\FTG}|^2 \Sphi+ \gamma \Sn}
  \end{align}
  where $|\FTG|^2 = \left(|\FTG_x|^2 + |\FTG_y|^2 \right)$
  and $\Sphi$ and  $\Sn$ are
  the spatial power-spectral-density (PSD) of the phase and noise. The noise is
  assumed white and uncorrelated, thus constant over all the
  frequencies whereas the phase PSD is given by
\begin{equation}
\Sphi = 0.49 r_0^{-5/3}\left\{(2\pi)^2\left[\kappa_x^2 +
      \kappa_y^2 + (1/L_0)^2\right]\right\}^{-11/6}, 
\end{equation}
in which 
$r_0$ and $L_0$ the coherence length of the turbulence and the outer-scale respectively.
  \begin{proof}
    Full derivation is provided in Appendix. 
    \lemmaend
  \end{proof}

  In passing, note that the term in the denominator of Eq. (\ref{eq:Phi_wiener})
  \begin{equation}
|{\FTG}|^2\Sphi = \Ss^\parallel ,
  \end{equation}
  is the PSD of the measurements within the pass-band. With the appropriate modifications the Wiener in Eq. (\ref{eq:Phi_wiener}) filter can be used with any forward model investigated by replacing the forward operators by their respective counterparts for other models.

  A further scalar
  factor $\gamma$ is introduced to properly
  weigh the priors term to account for other unknown system parameters, of which aliasing is one contributor. 
  The term $\Sn/\Sphi$ can be interpreted as the
  inverse of the signal-to-noise ratio (SNR). At frequencies
  where the signal is very strong relatively to the noise,
  $\Sn/\Sphi\approx 0$ and the Wiener filter becomes of
  the form of the LSQ filter. For
  frequencies with weak signal $\Sn/\Sphi\rightarrow
  \infty$ with output  $\widehat{\FTWF} \rightarrow 0$ \textit{i.e.} the zero estimate is
  adopted. By proper choice of
  $\Sn \neq \text{constant}\,\, \forall\,\,\boldsymbol{\kappa}$ some degree of aliasing compensation can be achieved
  by the reconstructor. It won't be further investigates here though.   

The analytical derivation of the MMSE filter with the aliasing term in Eq. (\ref{eq:S_FT_AliasTerm}) is provided next.

  \subsection{Anti-aliasing Wiener filtering}
At this point, it is now a natural subsequent step to derive the MMSE filter for the case where the forward model includes explicitly an aliasing term in Eq. (\ref{eq:S_FT_AliasTerm}). The minimization of Eq. (\ref{eq:WienerFilterMinimFunctional}) becomes
  \begin{align}\label{eq:Phi_wiener_antialias}
    \FTR_{\text{\tiny{AA}}} (\boldsymbol{\kappa}) &  = \frac{\FTG^* \Sphi}{\Ss^\parallel + \Ss^\perp + \gamma \Sn},
  \end{align}       
  indeed very similar to Eq.~(\ref{eq:Phi_wiener_antialias}) where
  \begin{align}
    \Ss^\perp = \sum_{\mathbf{m}\neq 0}\left|\FTG\left(\boldsymbol{\kappa} + \mathbf{m}/d \right)\right|^2 \Sphi \left(\boldsymbol{\kappa} + \mathbf{m}/d \right),
  \end{align}
is the aliased portion of the slopes.
  \begin{proof}
The derivation of the anti-aliasing Wiener filter follows closely that of the Wiener filter in Appendix. 

We now consider an extra aliasing term in the measurement equation given by
  \begin{equation}\label{eq:S_A_FT}
    \FTS\left(\boldsymbol{\kappa}\right) = \FTG \FTWF\left(\boldsymbol{\kappa}\right) + \widetilde{\boldsymbol{\eta}}\left(\boldsymbol{\kappa}\right) + \widetilde{\boldsymbol{\alpha}}\left(\boldsymbol{\kappa}\right)
  \end{equation}
  where $\widetilde{\boldsymbol{\alpha}}\left(\boldsymbol{\kappa}\right)$ is the term representing the aliasing measured at the WFS -- Eq. (\ref{eq:S_FT_AliasTerm}).\\
  
Next plug Eq. (\ref{eq:S_A_FT}) in Eq. (\ref{eq:WienerFilterMinimFunctional}) and develop the squared modulus. 
Assuming that spatial frequencies are statistically independent the cross terms vanish as before to yield
      \begin{align*}
            \langle \varepsilon^2 \rangle & = \langle |\FTWF - \FTR \FTS|^2 \rangle = \Sphi - (\FTR^\ast\FTG^\ast + \FTR\FTG)\Sphi + \FTR\FTR^\ast (|\FTG|^2\Sphi  + \Ss^\perp + \Sn)
    \end{align*}
where $\Ss^\perp = 
\sum_{\mathbf{m}\neq 0}|{\widetilde{\mathcal Q} }\left(\boldsymbol{\kappa} + \mathbf{m}/d \right)|^2 \Sphi \left(\boldsymbol{\kappa} + \mathbf{m}/d \right)$ is the slopes' PSD with the aliasing component.

To find the representation of $\FTR$ that minimizes the above equation we take derivatives with respect to the coefficients of $\FTR$ and equate to zero leading to
      \begin{align*}
            \frac{\partial \langle \varepsilon^2 \rangle}{\partial \FTR} = 0 & \Leftrightarrow -\FTG\Sphi + \FTR^\ast(|\FTG|^2\Sphi + \Ss^\perp + \Sn) = 0\\
                & \Leftrightarrow \FTR = \frac{\FTG^\ast \Sphi}{\underbrace{|\FTG|^2 \Sphi}_{\Ss^\perp} + \Ss^\perp + \Sn }. 
      \end{align*} \lemmaend
   
  \end{proof}

  The anti-aliased Wiener filter admits thus a very intuitive representation. The denominator of Eq. (\ref{eq:Phi_wiener_antialias}) is composed of the sum of three spatial PSDs: that of slopes with no aliasing $(\Ss^\parallel = | \FTG |^2\Sphi)$, the aliased component ($\Ss^\perp$) and the noise PSD $(\Sn)$ which added together represent the total PSD of the measurements $(\Ss)$. 

  \subsection{Fitting errors: in- and out-of-band}

In a standard AO system, the phase estimate is least-squares projected onto the bi-dimensional surface of a deformable mirror (DM) -- a deterministic optimization problem that due to the separation principle can be dealt with separately.
 
  The DM face-sheet shape is produced by a superposition of actuator
  influence functions $\mathcal{H}_{\text{\tiny{DM}}}(x,y)$ spatially separated by a pitch $d$ through the circular convolution
  \begin{equation}\label{eq:a_conv_IF}
    \WF^\text{cor}(x,y) = \sum_{m,n}  \mathcal{H}_{\text{\tiny{DM}}}(x-m d,y-n\,d)  \mathbf{u}(m\,d,n\,d),
  \end{equation}
  where $\mathbf{u}$ is the DM command map and $\WF^\text{cor}(x,y)$ is the high-resolution DM-produced correction phase. 




  The convolution of Eq. \eqref{eq:a_conv_IF} may be represented in  the
  Fourier domain by a multiplication, where the transform  $\widetilde{\mathbf{u}} \triangleq \TFd\left\{\mathbf{u}(m\,d,n\,d)\right\}$
  is periodic on an interval equal to $1/d$. Therefore controlling
  frequencies in the range $|\boldsymbol{\kappa}|<1/(2d)$ 
  will excite higher
  frequencies. 
  The transform $\FTWF^\text{cor}\left(\boldsymbol{\kappa}\right)$ is also periodic on an interval $1/d$. 
  Therefore within the correctable band one gets
  \begin{align}
    \widetilde{\mathcal{H}}_{\text{\tiny{DM}}} & = \sum_{\mathbf{m}} \widetilde{\mathcal{H}}(\boldsymbol{\kappa} - \mathbf{m}/d)\\
    & = \widetilde{\mathcal{H}}_0 + \sum_{\mathbf{m}\neq 0} \widetilde{\mathcal{H}}(\boldsymbol{\kappa} - \mathbf{m}/d) = \widetilde{\mathcal{H}}_0 +  \widetilde{\mathcal{H}}_\mathbf{m},
  \end{align}
where it is important to consider a DM filter that takes into account high-order terms beyond the correctable band that due to circularity fold into low-order, in-band components. 
  Such "anti-folding" DM fitting filter is 
  \begin{equation}
    \widetilde{\mathcal{F}}(\boldsymbol{\kappa}) = \frac{1}{\widetilde{\mathcal{H}}_0 +  \widetilde{\mathcal{H}}_\mathbf{m}} = \frac{1}{\widetilde{\mathcal{H}}_\text{\tiny{DM}}}
  \end{equation}

  \begin{proof}
We seek
\begin{align*}
    \mathbf{\widetilde{u}}_{\text{\tiny{DM}}} & = 
    \arg \min_{\mathbf{\widetilde{u}}} |\FTWF - \widetilde{\mathcal{H}}_{\text{\tiny{DM}}} \widetilde{\mathbf{u}}|^2 
  \end{align*}
where $\epsilon^2 = |\FTWF - \widetilde{\mathcal{H}}_{\text{\tiny{DM}}} \widetilde{\mathbf{u}}|^2  = |\FTWF - \FTWF_{\text{\tiny{DM}}}|^2$ with $\FTWF_{\text{\tiny{DM}}} =  \widetilde{\mathcal{H}}_{\text{\tiny{DM}}}\mathbf{\widetilde{u}}$. Then
  \begin{align*}
    \frac{\partial \epsilon^2}{\partial \mathbf{\widetilde{u}}}=0
   & \Leftrightarrow
   -\FTWF \widetilde{\mathcal{H}}_{\text{\tiny{DM}}}^\ast\frac{\partial\mathbf{\widetilde{u}}^\ast}{\partial\mathbf{\widetilde{u}}}
   -\FTWF^\ast_{\text{\tiny{PR}}} \widetilde{\mathcal{H}}_{\text{\tiny{DM}}}\frac{\partial\mathbf{\widetilde{u}}}{\partial\mathbf{\widetilde{u}}}
   +\widetilde{\mathcal{H}}_{\text{\tiny{DM}}}\widetilde{\mathcal{H}}_{\text{\tiny{DM}}}^\ast\frac{\partial\mathbf{\widetilde{u}}\mathbf{\widetilde{u}}^\ast}{\partial\mathbf{\widetilde{u}}}=0
\\   & \Leftrightarrow
   -\widetilde{\mathcal{H}}_{\text{\tiny{DM}}}\FTWF^\ast_{\text{\tiny{PR}}} 
   +\widetilde{\mathcal{H}}_{\text{\tiny{DM}}}\widetilde{\mathcal{H}}_{\text{\tiny{DM}}}^\ast\mathbf{\widetilde{u}}^\ast = 0
\\
   & \Leftrightarrow
   \mathbf{\widetilde{u}} = \frac{1}
   {\widetilde{\mathcal{H}}_{\text{\tiny{DM}}}}\FTWF 
     \end{align*}
    \lemmaend
  \end{proof}

  The DM commands, \textit{i.e.} the
  coefficients of the DM influence functions are
  obtained from
  \begin{equation}\label{eq:DMfilter}
    \widetilde{\mathbf{u}} = \widetilde{\mathcal{F}} \FTR \FTS = \frac{\widetilde{\mathcal{H}}_{\text{\tiny{DM}}}^*}{|\widetilde{\mathcal{H}}_{\text{\tiny{DM}}}|^2} 
\FTR \FTS ,
  \end{equation}
  with the  corrected WF 
  \begin{equation}
    \FTWF^\text{cor} = \widetilde{\mathcal{H}}_{\text{\tiny{DM}}} \widetilde{\mathbf{u}} =  \widetilde{\mathcal{H}}_{\text{\tiny{DM}}} \widetilde{\mathcal{F}} \FTR \FTS .
  \end{equation}
  where $\widetilde{\mathcal{F}} $ is a linear filter through which the least-squares fit to the DM influence functions is made.
In order to minimize the in-band fitting errors the commands need be computed with the full folded filter. Contrarily an error arises from only considering the in-band DM response. This error can  be assessed numerically integrating $\xi^2 = |1-\widetilde{\mathcal{H}}_{\text{\tiny{DM}}}/\widetilde{\mathcal{H}}_{0}|^2 \Sphi$ over the spatial frequencies. Since the term $ |1-\widetilde{\mathcal{H}}_{\text{\tiny{DM}}}/\widetilde{\mathcal{H}}_{0}|^2\neq \mathbf{0}$ quantitative estimates will reveal perceptible errors that can be further mitigated by proper filtering, i.e. choosing $\widetilde{\mathcal{F}}$ as in Eq. (\ref{eq:DMfilter}). 



\begin{figure}[htpb]
    \begin{center}
      \includegraphics[width=1.0\textwidth]{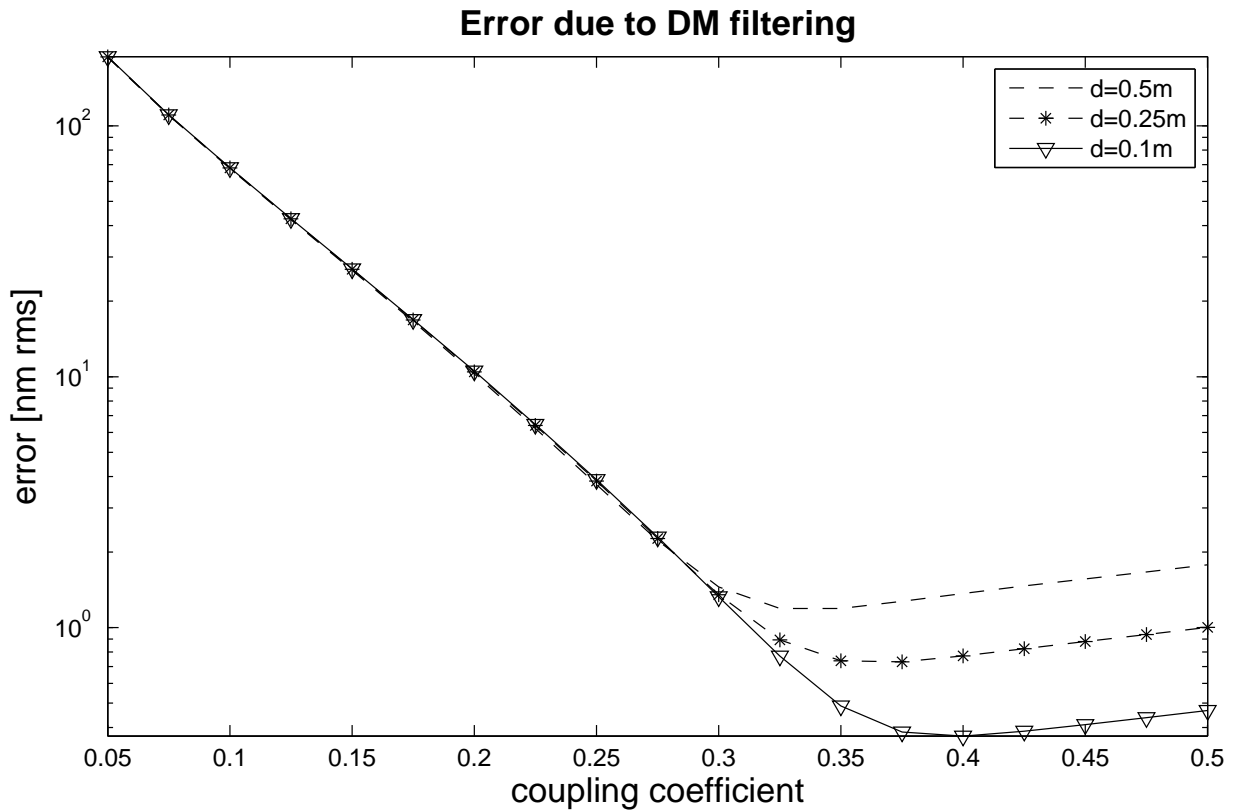}
    \end{center}
    \caption[]
    {\label{fig:fig_DMfiltering_Error_fcn_coupling}
      In-band DM fitting error in nm rms for a $r_0$=15\,cm and $L_0$=30\,m (piston-removed) with improper DM filtering for a Gaussian influence-function with varying coupling coefficient. The fitting error is a smooth function of the DM pitch. Minima are found for increasing coupling coefficients when decreasing DM pitches. These minima occur for coupling coefficients around [0.3-0.4], rather common values in AO. 
}
  \end{figure}

  Figure \ref{fig:fig_DMfiltering_Error_fcn_coupling} investigates the incremental fitting error within the correctable band as a function of the DM cross-coupling coefficient. Remarkably enough, values $\sim$30-40\% coupling provide minima: 
 for lower couplings the influence-functions are too spiky and originate a bad fit to low turbulence frequencies whereas for higher couplings the influence functions cannot fit the high-frequency content of the phase spectrum. This also agrees with Monte-Carlo simulations of DM fitting from randomly generated Kolmororov/von-K\'arm\'an phase screens.
In all, for the standard coupling values used in AO the impact of improper filtering can safely be neglected. The results in Fig. \ref{fig:fig_DMfiltering_Error_fcn_coupling} provide reassuring evidence to support that common procedure.

  The out-of-band fitting error is computed from 
\begin{equation}\label{eq:fitting_error}
\sigma^2_\text{fit}= \int_{|\boldsymbol{\kappa}|\geq(2d)^{-1}}  |1-\widetilde{\mathcal{H}}_{\text{\tiny{DM}}}|^2\Sphi(\boldsymbol{\kappa}) \dint \boldsymbol{\kappa} \approx \int_{|\boldsymbol{\kappa}|\geq(2d)^{-1}}\Sphi(\boldsymbol{\kappa}) \dint \boldsymbol{\kappa},
\end{equation}
where despite $\widetilde{\mathcal{H}}_{\text{\tiny{DM}}}\neq \mathbf{0}$ for $|\boldsymbol{\kappa}|\geq 1/(2d)$ for all reasonable purposes the product of $\widetilde{\mathcal{H}}_{\text{\tiny{DM}}}$ by the phase PSD which is $\Sphi(\boldsymbol{\kappa}) \propto \boldsymbol{\kappa}^{-11/3} $ is practically null.
  \section{Residual  wave-front error: reconstruction, aliasing and noise errors}\label{sec:WF_alias_noise_error}

 
The results derived so far provide suitable formulations for evaluating the individual error contributors for the global budget.

Let the piston-removed phase PSD be such that
\begin{equation}\label{eq:PRfilter}
\Sphi'(\boldsymbol{\kappa}) = \left[1-\left|\frac{2 J_1(\pi \boldsymbol{\kappa} D)}{\pi \boldsymbol{\kappa} D} \right|^2 \right] \Sphi(\boldsymbol{\kappa}) =  \widetilde{\mathcal{P}} \Sphi(\boldsymbol{\kappa})
\end{equation}
in which $J_1(\cdot)$ a Bessel function of the first kind to accomplish the computation of the Fourier transform of a circular pupil of diameter $D$. $\widetilde{\mathcal{P}}(\boldsymbol{\kappa})$ is the piston-removal operator within square brackets in Eq. (\ref{eq:PRfilter}). 
 
 Using the Parseval theorem, the residual phase variance (piston-removed) is defined by
  \begin{equation}\label{eq:reconstruction_error}
    \sigma^2_\text{Tot}(d,D,r_0,L_0,\sigma^2_\eta, T_s) \triangleq \int_{-(2d)^{-1}}^{(2d)^{-1}} 
     \widetilde{\mathcal{P}} \left\langle \left| \FTWF(\boldsymbol{\kappa}) - {\FTWF}^{\text{cor}}(\boldsymbol{\kappa})  \right|^2 \right\rangle
    \partial \boldsymbol{\kappa}
  \end{equation}
 In the remainder we suppose that the compensated DM filtering is taking place, \textit{i.e.} ${\FTWF}^\text{cor}(\boldsymbol{\kappa}) = \widehat{\FTWF}(\boldsymbol{\kappa})$ when the anti-folding filter is applied. The failing case is easily computed by updating the following equations with the results from the previous section.

Expanding Eq. (\ref{eq:reconstruction_error}) with $\widehat{\FTWF}(\boldsymbol{\kappa}) = \FTR\FTS(\boldsymbol{\kappa})$ and the measurement Eq. (\ref{eq:S_FT}) one gets
  \begin{equation}
     \widetilde{\mathcal{P}}\average{\left| \FTWF(\boldsymbol{\kappa}) - \widehat{\FTWF}(\boldsymbol{\kappa})  \right|^2}
    = \average{|\FTWF_\perp|^2} 
    + \left|1 - \FTR \FTG\right|^2 \widetilde{\mathcal{P}} \average{\FTWF(\boldsymbol{\kappa})\FTWF(\boldsymbol{\kappa})^* }
    + \mathbf{W}_\text{RA}
    + \average{ \widetilde{\mathcal{P}} \left|\FTR \FTN\right|^2} 
  \end{equation}
  with $
  \average{|\FTWF_\perp|^2}$ the integrand in the fitting error Eq. (\ref{eq:fitting_error}) where we considered that $ \widetilde{\mathcal{P}}(\boldsymbol{\kappa}) = 1$ for $|\boldsymbol{\kappa}|>1/(2d)$; the term 
\begin{equation}
\left|1 - \FTR \FTG\right|^2 \widetilde{\mathcal{P}} \average{\FTWF(\boldsymbol{\kappa})\FTWF(\boldsymbol{\kappa})^* }=  \left|1 - \FTR \FTG\right|^2\Sphi'(\boldsymbol{\kappa})
\end{equation}
 is  the static phase reconstruction error PSD, $\average{ \widetilde{\mathcal{P}} \left|\FTR \FTN\right|^2}$ is the noise propagated PSD and 
  \begin{equation}
    \mathbf{W}_\text{RA}
= \widetilde{\mathcal{P}} \sum_{\mathbf{m}\neq 0}  \left|\FTR(\boldsymbol{\kappa}) \FTG(\boldsymbol{\kappa}+ \mathbf{m}/d)\right|^2 \Sphi(\boldsymbol{\kappa}  + \mathbf{m}/d)
  \end{equation}
  the reconstructed aliasing PSD. 
  \subsection{Aliasing and fitting error as a function of $D/d$}

\begin{figure}[htpb]
    \begin{center}
      \includegraphics[width=1.0\textwidth]{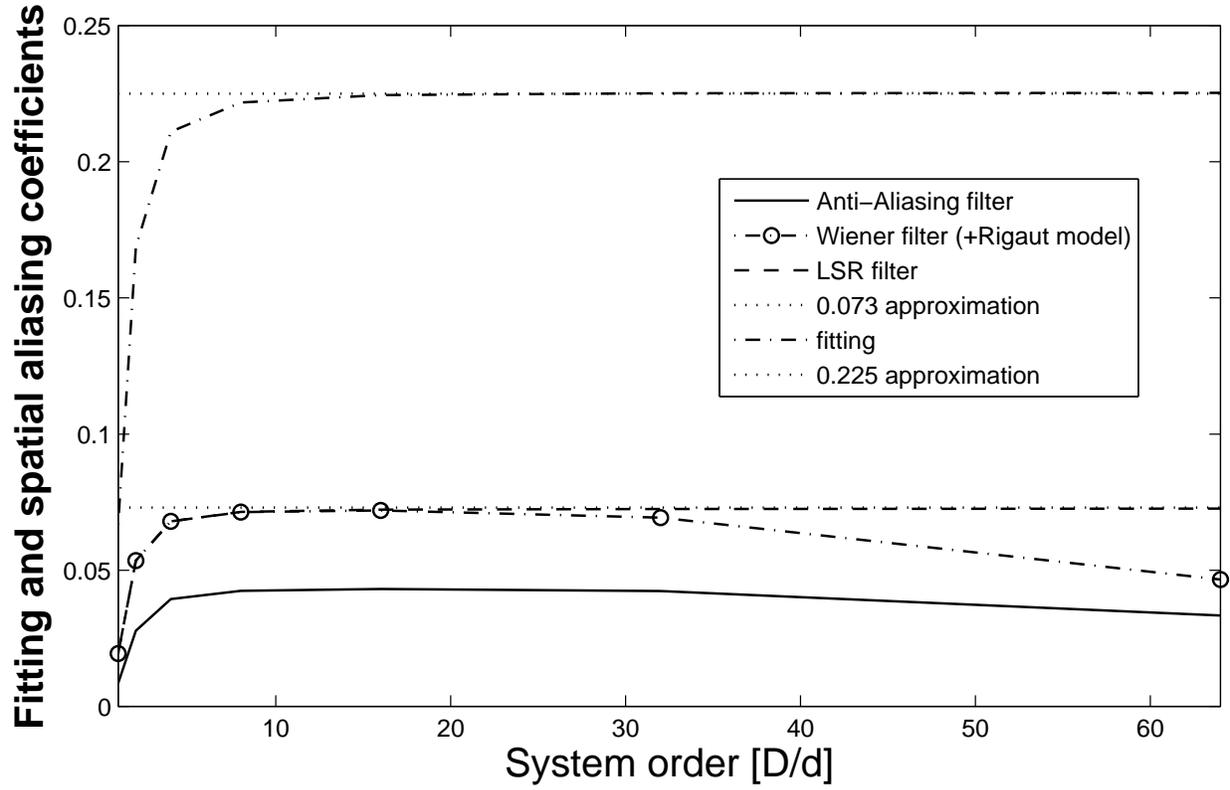}\\\includegraphics[width=1.0\textwidth]{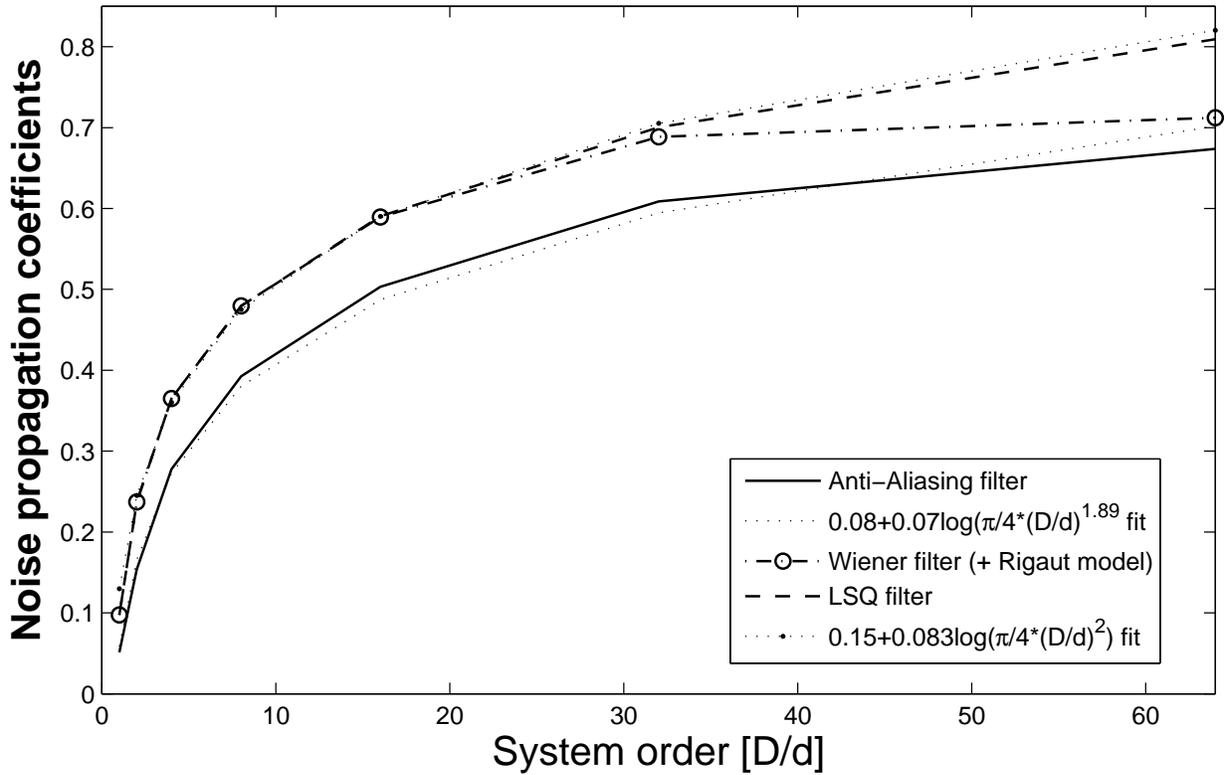}  
    \end{center}
    \caption[]
    {\label{fig:fig_AliasFit_MVAlias_AAAlias_propagation}
      Propagation errors as a function of the system order. The AA filter aliasing propagation decreases with system order to values $\approx$0.035 and $\approx$0.01 (32x32 and 64x64 respectively), considerably lower that the LSQ value of 0.073. Noise propagation is also considerably reduced by a factor of $\approx$ 30\% with respect to the LSQ filter.}
  \end{figure}

The aliasing and noise propagation are now be assessed as a function of the system order $D/d$. Fig. \ref{fig:fig_AliasFit_MVAlias_AAAlias_propagation} depicts such results for the LSQ, Wiener and Anti-Aliasing Wiener filters.  We have reproduced the results in \textit{Ellerbroek,05} \cite{ellerbroek05} and overlaid the new results for the aliasing and noise propagation following the parameters of Table \ref{tab:baseline}. We compute the out-of-band fitting error to be $0.225 (d/r_0)^{5/3}$ (slightly less than in \cite{ellerbroek05}) whereas the aliasing decreases from its LSQ value of $0.073(d/r_0)^{5/3}$ to $\approx 0.035(d/r_0)^{5/3}$ and $\approx 0.01(d/r_0)^{5/3}$ for the Anti-Aliasing Wiener filter (32x32 and 64x64 cases respectively); in terms of noise propagation, the AA filter has a fit with smaller growth rate: $log(\pi/4 (D/d)^(1.89))$ instead of$log(\pi/4 (D/d)^(2))$. The behavior of the Wiener is more complex though: for a system order below 20 it behaves like the LSQ filter but for higher order SCAO systems it converges to the AA filter. The same applies to noise propagation.

  \section{Sample numerical results for a 32X32 and 64x64 system}

The analytical filters derived in the previous sections will now be applied to a 32x32 and a 64x64 sub-aperture high-order AO system. This provides insight into the potential achievements of the new anti-aliasing Wiener filter for systems that cover the range of current HCI systems. Table \ref{tab:baseline} summarizes the baseline simulation parameters assumed in the results that follow.
  \begin{table}[h!]
    \caption
    {Baseline Configuration Parameters.}
    \vskip 2mm
    \begin{center}
      \begin{tabular}{ll}
        \hline \hline
        {\bf Telescope} & \\
        D & 8 m \\
        \hline
        {\bf Atmosphere} & \\
        $r_0$ &  15\,cm \\
        $L_0$ & 30\,m \\
        zenith angle & 0\,deg \\
        Altitudes & 0\,km\\
        wind speed & 10\,m/s \\
        wind direction & 0\,deg \\
        Sampling & 8/400\,m \\
        \hline
        {\bf Shack-Hartmann Sensor} & \\
        NGS V magnitude & 10 \\
        Readout Noise (RON) & 0\,e$^-$ \\
        Order & [32$\times$32; 64$\times$64]\\
        Width $d$& [0.25, 0.125]\,m\\
        $N_{pix}$ & 4 (linear)\\
        $f_{sample}$ & 1000\,Hz \\
        $\lambda_{WFS}$ & 0.55\,$\mu$m \\
        \hline
        {\bf DM} & \\
        pitch & [8/32, 8/64]\,m \\
        influence & Gaussian \\
        coupling & 30\%\\
        \hline
        \hline
      \end{tabular}
    \end{center}
    \label{tab:baseline}
  \end{table}




  \subsection{Error breakdown using PSD analysis}

We may also explore and visually inspect results provided by the analytical formulation developed in this paper to depict bi-dimensional PSDs for the phase reconstruction, aliasing and noise propagated. A detailed error breakdown using numerical integration of the PSDs over frequencies within the correctable band is utilized to quantitatively assess and compare performance -- Table \ref{tab:32x32} summarizes results for a $32\times32$ and $64\times 64$ AO system on a 8\,m telescope.

Predicted performance shows that the filters based on the Rigaut model and the AA filter do actually increase the overall wave-front error for a $64\times 64$ with respect to a smaller $32\times32$ system. By inspecting the table  \ref{tab:32x32} noise propagation is the main culprit.  
\begin{table}[h!]
    \caption
    {Error breakdown for a 32x32 (top) and 64x64 (bottom) system. Values quoted in nm\,rms. Second to last row shows the incremental error (in quadrature) with respect to the AA Wiener filter. Last row quotes the Strehl-ratio in J-band (1.65\,$\mu$m). 
}
    \vskip 2mm
\begin{center}
\begin{tabular}{|c|c|c|c|c|c|}
  \hline
   & Fried & Hudgin & Southwell & Rigaut & AA  \\
  \hline
  $\sigma_\varphi   $ & 49.41 & 55.59 & 59.08 &  2.28 & 20.09 \\
  \hline
  $\sigma_\eta   $ & 22.25 & 35.13 & 15.41 & 35.19 & 27.52 \\
  \hline
  $\sigma_\alpha $ & 37.80 & 21.48 & 18.56 & 22.14 & 20.82 \\
  \hline
  $\sigma_\text{Tot}$ & 66.08 & 69.18 & 63.82 & 41.64 & 39.93 \\
  \hline
  Incr Error         & 52.64 & 56.49 & 49.79 & 11.81 &  \\
  \hline
  Strehl@1.65$\mu$m & 0.969 &   0.967  &   0.971 &   0.988 &   0.989\\
\hline
\end{tabular}
 \end{center}
 \begin{center}
\begin{tabular}{|c|c|c|c|c|c|}
  \hline
   & Fried & Hudgin & Southwell & Rigaut & AA  \\
  \hline
  $\sigma_\varphi   $ &  27.36 & 34.53 & 37.32 & 10.93 & 16.75 \\
  \hline
  $\sigma_\eta   $ & 44.76 & 44.51 & 40.36 & 45.03 & 43.81 \\
  \hline
  $\sigma_\alpha $ & 16.38 & 16.44 &  7.96 & 16.10 & 13.63 \\
  \hline
  $\sigma_\text{Tot}$ & 54.94 & 58.68 & 55.54 & 49.05 & 48.84 \\
  \hline
  Incr Error         & 25.14 & 32.53 & 26.45 &  4.55 &  \\
  \hline
Strehl@1.65$\mu$m &  0.979 &   0.977  &  0.979  &  0.984  &  0.984 \\
\hline
\end{tabular}
\end{center}
    \label{tab:32x32}
  \end{table}

\begin{figure}[htpb]
    \begin{center}
      \includegraphics[width=1.0\textwidth]{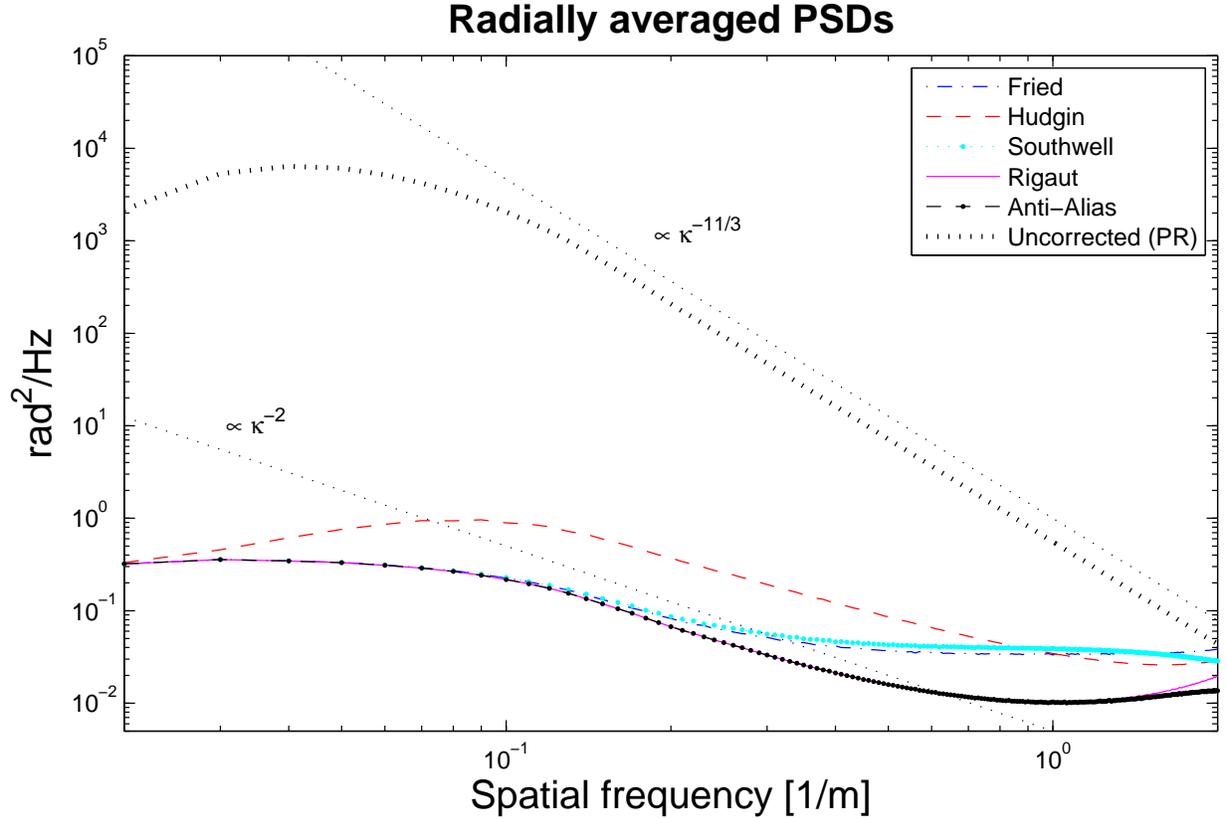}
    \end{center}
    \caption[]
    {\label{fig:PSDs_ERRTot}
      Residual wave-front PSDs. The characteristic  wave-front $\propto \boldsymbol{\kappa}^{-11/3}$ and noise propagation $\propto \boldsymbol{\kappa}^{-2}$ are indicated in black-dotted lines in the figure. The positive slope of the uncorrected phase PSD at low spatial frequencies is due to the piston-removal filtering function.}
  \end{figure}

Figure \ref{fig:PSDs_ERRTot} shows the radially averaged PSD cuts where one can visualize the location of the errors and departures from the $\boldsymbol{\kappa}^{-2}$ characteristic of noise propagation when both phase reconstruction error and aliasing are jointly considered.

This aggregated result shows the AA filter performing only slightly better than the Wiener filter. Both bear sensible improvements with respect to the approximate models in particular in the high-frequency end. 

\begin{figure}[!tp]
      \includegraphics[width=1.0\textwidth]{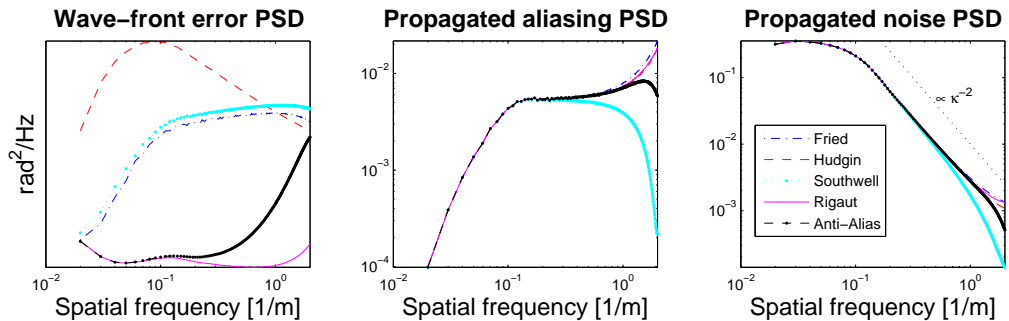}
    \caption[]
    {\label{fig:PSDs_ERRPhi_Alias_Noise}
      Radially-averaged PSD cuts. Is is apparent that both the Hudgin and Southwell models suffer mainly from a large phase reconstruction error which contributes the most for the overall error obtained. The Southwell makes for a rather curious case in that it is the best regarding aliasing and noise propagation with increased high-frequency rejection near the band-pass limit. Note the $\propto \boldsymbol{\kappa}^{-2}$ behavior of the noise propagation}
  \end{figure}

  Figure \ref{fig:PSDs_ERRPhi_Alias_Noise} splits the PSD into phase reconstruction error, aliasing and noise propagation - the fitting error is shared by all of them and thus not shown here. 

Analysis of these findings deserves several comments:
  \begin{itemize}
  \item Although the AA filter ensures the least wave-front residual variance -- cf. Fig. \ref{fig:PSDs_ERRTot} -- it is the Southwell filter (with optimal shifts) that propagates the least noise. This is in perfect agreement with the results of \cite{zou06} for the Southwell model and also for the  Fried and Hudgin models. 
  \item Remarkably, regarding aliasing propagation it is the Southwell model that propagates the least and not the AA filter as one would intuitively expect. However despite lower noise and aliasing propagation such result is  over-weighted by a greater phase reconstruction error for a large range of frequencies within the pass-band.
  \item  From these results the Hudgin model, albeit with optimal extra alignments, bears the worst if only slightly residuals for both the  $32\times32$ and $64\times 64$ systems sizes.
  \end{itemize}




\subsection{Regularization weighting factors adjustment}
 
\begin{figure}[!tpb]
\begin{center}
\includegraphics[scale=1.0]{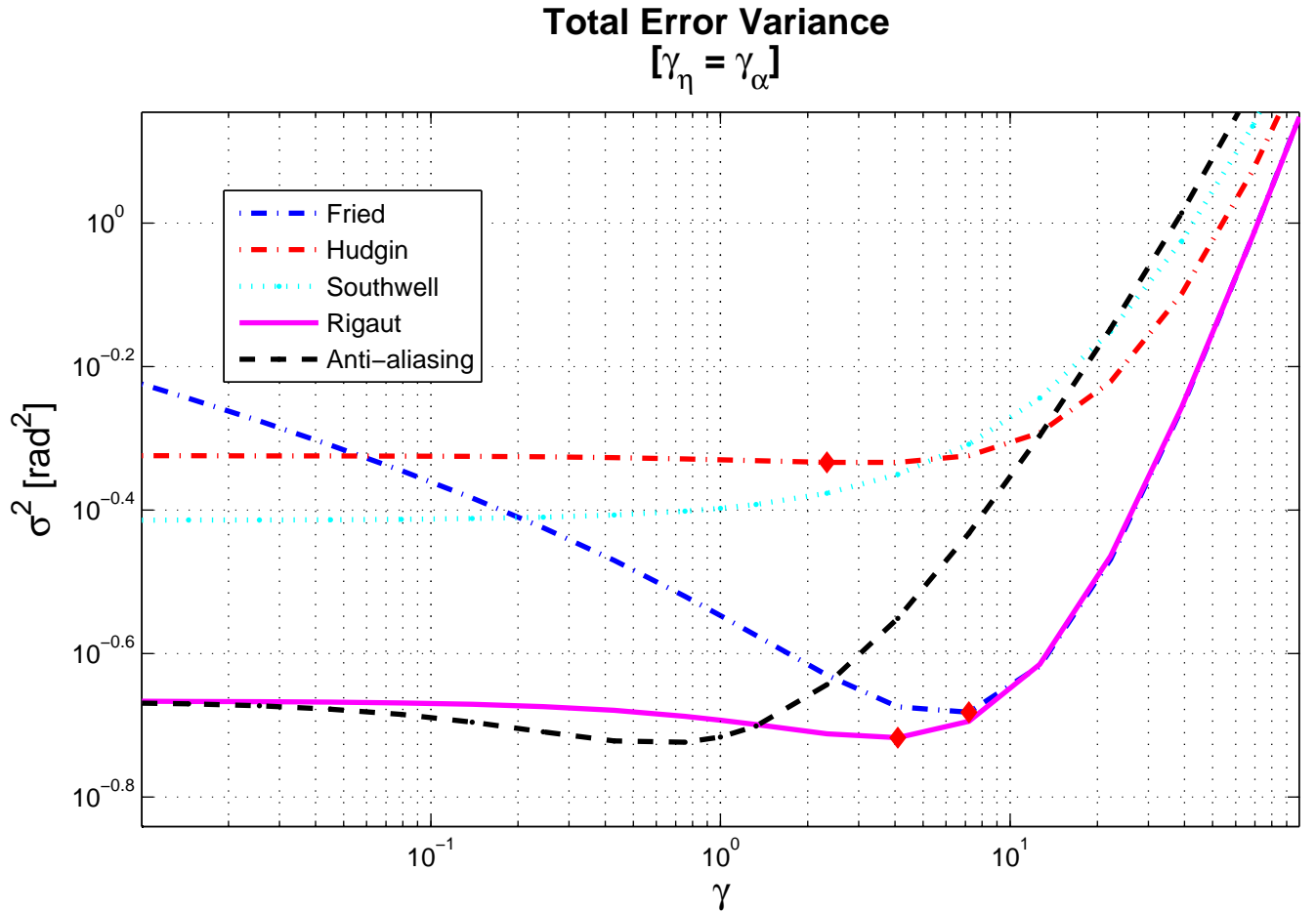}
\end{center}
\caption[]
    {\label{fig:gamma_performance_shift_total}
     Total residual error as a function of the regularizing parameter $\gamma$. By boosting the regularization to compensate for aliasing not modeled the Wiener filter can achieve roughly the same overall error as the AA Wiener filter.}
\end{figure}

It is instructive to explore the effect of over-regularization in the Wiener filters. 
Figure \ref{fig:gamma_performance_shift_total} shows the variation of residual wave-front error variance from Eq. (\ref{eq:reconstruction_error}) as a function of the regularization parameter in Eq. (\ref{eq:Phi_wiener}). The AA filter achieves the best performance of all the filters and that for a $\gamma=1$, as expected. Figure \ref{fig:gamma_performance_shift_total} also shows that parameter $\gamma$ needs be slightly boosted to accommodate the aliasing error term that is not taken into account explicitly in the reconstructor. 
For either the Rigaut, Fried and Hudgin filters a value roughly of $\gamma\approx 10$ is found whereas for the Southwell filter the impact of the regularization is not perceptible if at all with the LSQ filter providing the best result. 

Note also that the Wiener filter minimum error with $\gamma \sim 10$ is about the same as the AA filter with $\gamma=1$, \textit{i.e.} over-regularizing the Wiener filter has the global effect of practically mitigating all the aliasing component.
 
  \subsection{High-contrast imaging systems: PSF morphology and achievable raw contrast}

  For high-contrast imaging wave-front error assessment is of prime importance as well as achievable contrast. In this section the raw PSF intensity is computed to give insight over the repartition of contrast as a function of angular separation and provide a comprehensive comparison between filters. Contrast is assumed as the ratio of the PSF raw intensity to its maximum value. 

The PSFs are computed from the PSDs from well-established relationships as follows. We compute covariance functions from the PSDs using the Wiener-Khinchine theorem \cite{flicker07a, rigaut98}. With those the spatial structure function can be seamlessly computed from which the Optical Transfer Function (OTF) is determined. The PSF is found from the Fourier transform of the OTF.

\begin{figure}[!tpb]
    \begin{center}
      \includegraphics[width=1.0\textwidth]{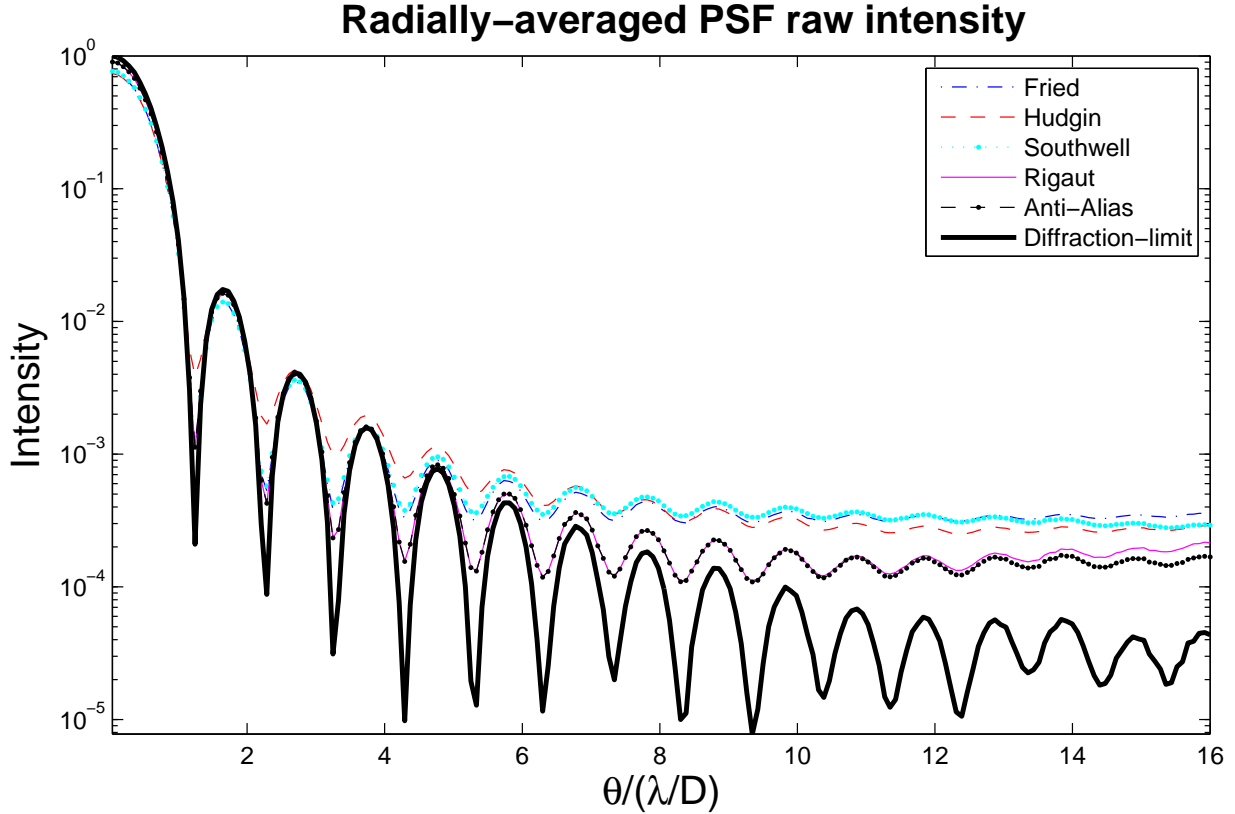}
    \end{center}
    \caption[]
    {\label{fig:fig_RAW_CONTRAST_CUT}
      Radially-averaged PSF raw intensities at the SH-WFS wavelength. 
      Strehl-ratios obtained in J band (1.65$\mu$m) are  [0.969; 0.967; 0.971; 0.988; 0.989] respectively.}
  \end{figure}

  Figure \ref{fig:fig_RAW_CONTRAST_CUT} shows the radially averaged cross-sections of the PSFs. The Wiener filters computed from the Rigaut SH-WFS models  bear consistent improved  raw contrasts of about a factor of 2 over the PSFs using the approximate phase-difference discrete models from Sect. \ref{sec:approx_disc_meas_models}. The AA filter does improve slightly the raw contrast for separations $\theta \geq 10 \lambda/D$. As shown in Fig. \ref{fig:PSDs_ERRTot} the Hudgin model is particularly affected by errors in the mid-range spatial frequencies which is also visible in Fig. \ref{fig:fig_RAW_CONTRAST_CUT} between $2$ and $10\,\lambda/D$.

\begin{figure}[htpb]
    \begin{center}
      \includegraphics[width=1.0\textwidth]{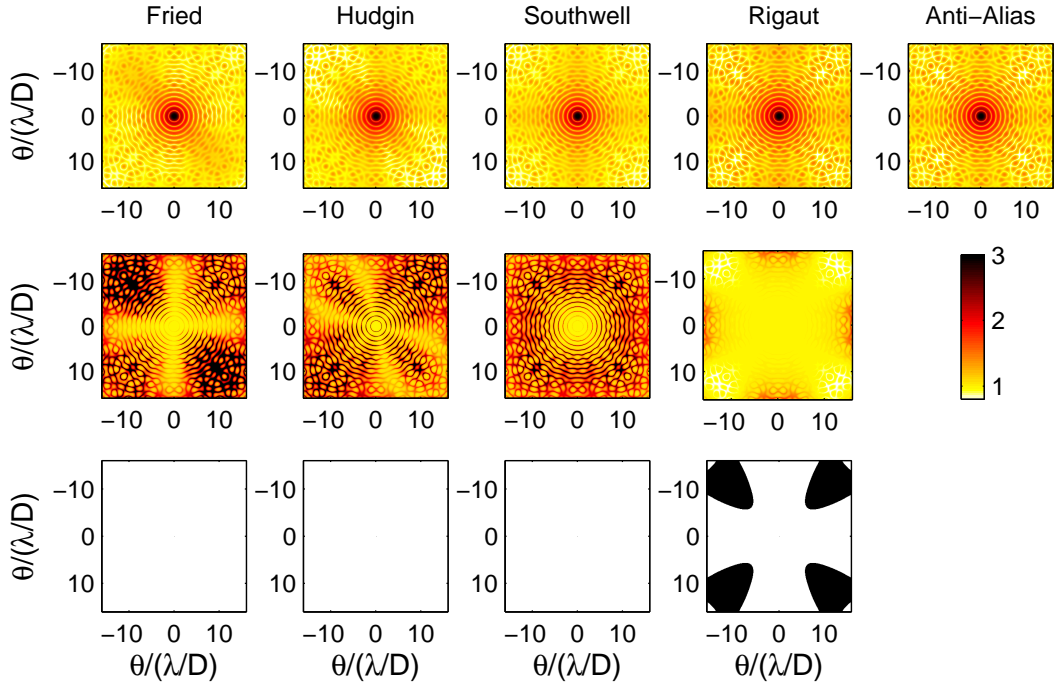}
    \end{center}
    \caption[]
    {\label{fig:fig_FACEON_PSF_COMPARISON}
      (Top) Face-on PSF normalized raw intensities in log scale at $\lambda = 1.65\,\mu m$. (Middle) point-wise ratio of contrasts achievable, where contrast is computed from the raw PSF intensities divided by the peak intensity; results depicted on a scale [0.8 and 3]. (Bottom) binary mask applied to the second row: within black regions the contrast is higher than the Anti-Alias filter contrast and otherwise for the white regions \textit{i.e.} the AA achieves best contrast. The AA average contrast is $1.9,1.7, 1.6$ better than those obtained with the filters based on the Fried, Hudgin and Southwell models respectively.}
  \end{figure}

  The face-on pattern of the in-band PSF (log scale) is depicted in Fig. \ref{fig:fig_FACEON_PSF_COMPARISON}.  The bottom row is a binary mask for intensities higher than the AA filter (blue) and lower that the AA (white). The AA raw contrast is higher than that of Fried's, Hudgin's and Southwell's all across the correctable separations. Regarding the Wiener filter it excels in the north-south and east-west lobes but not on the diagonal ones where the Wiener achieves better contrast. The AA average contrast still is better ($\sim$4\%) than the Wiener's with contrasts that can reach 1.7 better and never less than 0.8 of the Wiener's. 

  \section{Summary}
This paper provides formulae to compute the Wiener filter and the optimal anti-aliasing Wiener filter in the spatial frequency domain for AO systems employing the Shack-Hartmann WFS. We have extensively compared both these solutions to approximate discrete models for the SH-WFS (commonly called Fried, Hudgin and Southwell) in terms of overall performance (Strehl-ratio) and PSF raw intensities.

  We have shown how to optimally deal with the periodicity  during the projection onto general DM influence functions and quantified the loss in performance of not taking into account high-order terms beyond the pass-band that fold in the correctable region. This error is almost negligible (sub-nanometer) for coupling coefficients in the 25\%-40\% range and grows rapidly when one departs from that.

  We have provided new wave-front error, noise propagation and aliasing estimates for the optimal anti-aliased filter with a new noise propagation that is about $70\%$ of the least-squares estimate and a revised aliasing of $0.035(d/r_0)^{5/3}$ instead of $0.073(d/r_0)^{5/3}$, for a 32x32 system with a magnitude 10 star (aliasing propagation is now a function of the measurement noise through the regularising term in the reconstructor -- see Table \ref{tab:baseline}).

  Regarding HCI specifically, we've shown pre-coronographic average raw contrast ratios achievable, reaching factors of up to 2 in comparison with the approximate models. Albeit, the anti-aliased filter fails to provide meaningful improvement over the straight Wiener filter. Further simulations tailored to HCI instruments need now be run to fully assess effectiveness and achievable gains. This results also stress a fundamental limit imposed on recovering information from noise-degraded signals. Such limits do indeed justify the deployment of AO systems to overcome detrimental SNR regimes and partially recover AO-corrected PSFs using before-the-fact phase-conjugation. 

  The results herein will now be extended to include a coronographic image formation model and generalized to the closed-loop case. Monte-Carlo simulations of high-order AO systems will ensue where filters need be discretized. In these cases the extension of the measurements outside the telescope's aperture may become a major factor and eventually modify the main findings of this paper.

  \section*{Acknowledgments}
  C. Correia acknowledges the support of the European Research Council through the Marie Curie Intra-European Fellowship with reference FP7-PEOPLE-2011-IEF, number 300162. All the simulations and analysis done with the object- 
  oriented MALTAB-based AO simulator (OOMAO) \cite{conan14} freely available from \htmladdnormallink{https://github.com/rconan/OOMAO/}{https://github.com/rconan/OOMAO/}

  \appendix
  \section{Wiener filter derivation}
We wish to find the filter $\FTR$ that minimizes the residual phase variance $\average{ \varepsilon^2(\boldsymbol{\kappa}) }$ in the correctable band, assuming the signal and noise processes are second-order stationary and the slopes are given by Eq. (\ref{eq:S_FT})
      \begin{align*}
            \average{ \varepsilon^2 } & = \average{ |\FTWF - \FTR \FTS|^2 } = \average{ (\FTWF - \FTR\FTS)(\FTWF - \FTR\FTS)^\ast }\\
                & = \average{ \FTWF \FTWF^\ast - \FTWF \FTR^\ast \FTS^\ast - \FTWF^\ast \FTR \FTS + \FTR \FTR^\ast \FTS \FTS^\ast } \\
                & = \average{ |\FTWF |^2 } - \FTR^\ast\average{\FTWF \FTS^\ast } - \FTR \average{ \FTWF^\ast \FTS } + \FTR \FTR^\ast \average{ \FTS \FTS^\ast }\\
                & =\Sphi -\FTR^\ast \FTG^\ast \Sphi -\FTR^\ast\average{ \FTWF\FTN^\ast } -\FTR \FTG \Sphi -\FTR\average{ \FTWF^\ast\FTN } \\ & \hspace{20pt} + \FTR \FTR^\ast( |\FTG|^2 \Sphi
                + \FTG^\ast\average{\FTN\FTWF^\ast } + \FTG \average{ \FTN^\ast\FTWF} + \Sn )\\
    \end{align*}
    
Since the signal and noise processes are independent, the expected value of the joint process is equal to zero. This yields
$$ \average{ \varepsilon^2 } = (1-\FTR^\ast\FTG^\ast - \FTR\FTG)\Sphi + \FTR\FTR^\ast (|\FTG|^2\Sphi + \Sn)$$



To find the representation of $\FTR$ that minimizes the above equation we compute the partial derivatives with respect to $\FTR$ and equate to zero 
      \begin{align*}
            \frac{\partial \average{ \varepsilon^2 }}{\partial \FTR} = 0 
            & \Leftrightarrow \left(1- \frac{\partial \FTR^\ast}{\partial \FTR}\FTG^\ast 
                                    - \frac{\partial \FTR}{\partial \FTR}\FTG\right)\Sphi + 
                                    \frac{\partial \FTR\FTR^\ast}{\partial \FTR}(|\FTG|^2\Sphi + \Sn) = 0\\
                & \Leftrightarrow -\FTG\Sphi + \FTR^\ast(|\FTG|^2\Sphi + \Sn) = 0\\
                & \Leftrightarrow \FTR = \frac{\FTG^\ast \Sphi}{\underbrace{|\FTG|^2 \Sphi}_{\Ss^\perp} + \Sn}
      \end{align*}

  \bibliographystyle{apalike}
  {\bibliography{references}}

\end{document}